\documentstyle[11pt]{article}
\begin{document}
\baselineskip=22.0truept

{\large \bf \begin{center}
S matrix of collective field theory
\end{center}}
\vspace*{1 cm}

\begin{center} Dae-Yup Song\\
Department of Physics, Brown University, Providence, Rhode Island 02912\\
and\\
Department of Physics,\\
Sunchon National University, Sunchon 540-742, Korea\footnote{permanent address}
\end{center}
\vspace*{1 cm}
By applying the Lehmann-Symanzik-Zimmermann (LSZ) reduction formalism, we study
the S matrix of collective field theory in which fermi energy is larger than
the height of potential. We consider the spatially symmetric and antisymmetric
boundary conditions. The difference is that S matrices are proportional to
momenta of external particles in antisymmetric boundary condition, while they
are proportional to energies in symmetric boundary condition. To the order of
$g_{st}^2$, we find simple formulas for the S matrix of general potential.
As an application, we calculate the S matrix of a case which has been
conjectured
to describe a "naked singularity".\newline
August 1994

\newpage

\begin{center} {\bf I. INTRODUCTION}\end{center}

The one-dimensional hermitian matrix model can be studied by fermionic method
and also by bosonic collective field theory. In the double scaling limit, the
fermionic methods are indeed powerful to give any scattering amplitude of
fermion density perturbation in terms of integrals of reflection
coefficients[1,2]. For some cases, as a computational tool, the fermionic
method
may be much more efficient than the bosonic one. As will be argued later,
however, the bosonic method[3] has its own right when complicated potential is
considered.

In Ref.[4], through a beautiful interpretation of an excitation of fermi
surface, the general form of classical collective field was given, and it was
used to make relation between incoming and outgoing waves which gives the S
matrix. The relation was completely solved to obtain the classical S matrix[5].
The study of the quantum collective field theory has been carried out[6,7], and
the
results reasonably agree with those from fermionic method or classical
method[1,2,4,5].

In the double-scaled quantum collective field theory of fermi energy smaller
than the height of potential (the $\mu>0$ case), divergences appears due to the
classical turning point[6], while there is no such divergence  for the $\mu<0$
case[8,7]. In the $\mu<0$ case, the "time of flight" $\tau$ could be defined
in $(-\infty,\infty)$ and the theory is described by a massless (Tachyon) field
 $X(t,\tau)$  in two-dimensional Minkowski spacetime. Since  two-dimensional
massless field X is unphysical and is a fluctuations  around some background,
it may be necessary to consider only the half of degree of freedom, as shown in
the Liouville theory[9]. This can be done by imposing the spatially symmetric
boundary condition  $(X(t,\tau)=X(t,-\tau))$ or spatially antisymmetric
boundary condition  $(X(t,\tau)=-X(t,-\tau))$. Since the quantum states
recognize global
information, in general quantum theory could be sensitive to the boundary
conditions[10].

In this paper, we will study the quantum S matrix of double-scaled $\mu<0$
collective field theory of {\em arbitrary potential}, by applying the LSZ
reduction method[11]. Both of the boundary conditions will be considered. For
the symmetric boundary condition it will be shown that S matrix can be written
in terms of
energies $(|p_i|)$ of external particles, while S matrix are proportional to
the momentum $(p_i)$  in the antisymmetric boundary condition. For the usual
inverted harmonic oscillator potential, our results of symmetric boundary
condition are basically same to those of Ref.[7] except  factors: square root
of product of external energies which are needed for the correct aymptotics.
However, a step further has been done in this paper; By making use of
convolution theorem, the S matrix of order of $g_{st}^2$ are written in terms
of velocity, which could simplify the analyses. As an application, the S matrix
of the potential  which has been conjectured to describe a "naked singularity"
are obtained[12,13].

In the next section, we introduce  Hamiltonian  and boundary condition. A brief
argument of applicability of LSZ reduction method in our case will be included,
since there has been no discussion in previous publications. Sec. III and
Sec.IV will be devoted to find the S matrices in symmetric and antisymmetric
boundary conditions, respectively. Applications of the general formulas will be
made in Sec.V. We conclude in Sec.VI with some remarks.

\begin{center}{\bf II. HAMILTONIAN AND BOUNDARY CONDITION}\end{center}

The Hamiltonian of $\mu <0$ collective field theory may be written as[7,8]
\begin{equation}
{\cal H}=-\frac{1}{4}\int_{-\infty}^\infty d\tau:[P^2+{X'}^2
    -\frac{\sqrt{\pi}}{\beta v^2}(PX'P+\frac{1}{3}{X'}^3)
    -\frac{1}{2\beta\sqrt{\pi}}(\frac{v''}{3v^3}-\frac{(v')^2}{2v^4})X']:,
\end{equation}
where the primes denote derivatives with respect to $\tau$ and the spatial
coordinate $\tau$ is related to the eigenvalue $\lambda$ of matrix model by
\begin{equation}
v(\lambda)=\frac{d\lambda}{d\tau}=\sqrt{2(\mu_F-U(\lambda))}.
\end{equation}

In the double-scaled theory, the velocity $v$ is expected to be proportional
to $\sqrt{|\mu|}=\sqrt{|\mu_c-\mu_F|}$ where $\mu_F$ and $\mu_c$ are fermi
energy and height of the potential $U(\lambda)$, respectively. And the coupling
constant of the theory  is $g_{st}=\frac{1}{\beta\mid \mu\mid}$. For
convenience, we define
$f_1(\tau), ~f_3(\tau)$ as
\begin{eqnarray}
f_1(\tau)&=&\mid \mu\mid  (\frac{v''}{3v^3}-\frac{(v')^2}{2v^4}),\\
f_3(\tau)&=&\mid \mu\mid \frac{1}{v^2(\tau)},
\end{eqnarray}
 and their Fourier transforms as
\begin{equation}
g_a(k)=\int_{-\infty}^\infty  e^{ik\tau}f_a(\tau)d\tau~~~~~~
\mbox{(for~~ any~~$a$)}.
\end{equation}
A very useful property of ${\cal H}$ is that[14,4]
\begin{equation}
{\cal  H}=\frac{1}{2}({\cal H}_R + {\cal H}_L),
\end{equation}
where
\begin{eqnarray}
{ \cal H}_R&=&\frac{1}{4}\int_{-\infty}^{\infty}d\tau
            :[(P-X')^2+\frac{\sqrt{\pi}}{3}g_{st}f_3(\tau)(P-X')^3
              +\frac{g_{st}}{2\sqrt{\pi}}f_1(\tau)(P-X')]:, ~~~~~   \\
{ \cal H}_L&=&\frac{1}{4}\int_{-\infty}^{\infty}d\tau
            :[(P+X')^2-\frac{\sqrt{\pi}}{3}g_{st}f_3(\tau)(P+X')^3
              -\frac{g_{st}}{2\sqrt{\pi}}f_1(\tau)(P+X')]: .~~~~~~
\end{eqnarray}

A two-dimensional massless field (tachyon) may be written as
\begin{equation}
X(t,\tau)=\int_{-\infty}^\infty\frac{dk}{\sqrt{4\pi\mid k\mid}}
      [a(k)e^{ik\tau-i\mid k\mid t}+a^\dagger(k)e^{-ik\tau+i\mid k\mid t}]
\end{equation}
with hermitian operators $a,~a^\dagger$ satisfying
\begin{equation}
[a(k),a^\dagger(k')]=\delta(k-k'),~~~~~[a(k),a(k')]=0.
\end{equation}
If we assume the symmetric boundary condition for $X$, the operator $a(k)$
satisfy
$a(k)=a(-k)$, which gives the relation
\[ (p+X')(t,\tau)=(P-X')(t,-\tau).\]
Therefore, if we impose the symmetric boundary condition on $X$, the
interaction term of
${\cal H}$ vanish and it becomes free theory.

In the antisymmetric boundary condition, the operator $a(k)$ satisfy
$a(k)=-a(-k)$ so that
$ (p+X')(t,\tau)=-(P-X')(t,-\tau)$. In this boundary condition, therefore,
\begin{equation}
{\cal  H}={ \cal H}_R={ \cal H}_L
\end{equation}
or
\begin{eqnarray}
{\cal H}&=&\int_0^\infty dk k:a^\dagger(k)a(k):\nonumber\\
  &&+\frac{ig_{st}}{12\pi}\int_0^\infty
     dk_1dk_2dk_3\sqrt{k_1k_2k_3}[g_3(k_1+k_2+k_3):a(k_1)a(k_2)a(k_3):
\nonumber\\
 &&~~~~~~~~~~~ -3g_3(k_1+k_2-k_3):a(k_1)a(k_2)a^\dagger(k_3):\nonumber\\
 &&~~~~~~~~~~~
+3g_3(-k_1-k_2+k_3):a^\dagger(k_1)a^\dagger(k_2)a(k_3):\nonumber\\
&&~~~~~~~~~~~
    -g_3(-k_1-k_2-k_3):a^\dagger(k_1)a^\dagger(k_2)a^\dagger(k_3):]\nonumber\\
&&-\frac{ig_{st}}{8\pi}\int_0^\infty
dk\sqrt{k}(g_1(k)a(k)-g_1(-k)a^\dagger(k)),
\end{eqnarray}
which will be considered as the Hamiltonian of our system afterwards.
A useful expression of ${\cal H}$ for practical calculation is
\begin{eqnarray}
{\cal H}&=&-\frac{1}{4}\int_{-\infty}^\infty d\tau :[P^2+X'^2]:\nonumber\\
       &&+\frac{\sqrt{\pi}g_{st}}{12}\int_{-\infty}^\infty f_3(\tau)
                   :[P_-(\tau)]^3:
  +\frac{g_{st}}{8\sqrt{\pi}}\int_{-\infty}^\infty f_1(\tau)P_-(\tau)~~~
\end{eqnarray}
where
\begin{equation}
P_-(\tau)\equiv -i\int_0^\infty\frac{dk}{\sqrt{\pi}}\sqrt{k}
    [a(k)e^{ik\tau-i\mid k\mid t}-a^\dagger(k)e^{-ik\tau+i\mid k\mid t}].
\end{equation}

Although the antisymmetric boundary condition is used to obtain the Hamiltonian
(12), we will
consider both of  symmetric and antisymmetric boundary conditions.  Actually
the S matix we
will evaluate is more similar to the previous result[4,5,6,7] in symmetric
boundary condition

The LSZ reduction method is to obtain $S_{\beta \alpha}=<\beta \mbox{out}\mid
\alpha \mbox{in}>=<\beta\mbox{in}\mid S\mid \alpha\mbox{in}>$  from the
time-ordered products of quantum fields[11]. The $\mid\alpha\mbox{in}>$ and
$<\beta\mbox{out}\mid$ are in- and out-states defined when interactions are
turned off. Though Hamiltonian of our system is not invariant under Lorentz
transformations, the free part (${\cal H}(g_{st}=0)$) is invariant under the
transformations, which enables us to define the in- and out-states.
Furthermore,
it is not difficult to find out that the following LSZ reduction formula is
true
for both of symmetric and antisymmetric boundary condition:
\begin{eqnarray}
&&<p_1p_2\cdots p_n\mbox{out}\mid q_1\cdots q_m \mbox{in}>\nonumber\\
&&~~=(\frac{i}{\sqrt{Z}})^{m+n}\prod_{i=1}^m\int dt_id\tau_i\prod_{j=1}^n\int
dt_j'd\tau_j' e^{iq_i\tau_i-i\mid q_i\mid t_i} e^{-ip_j\tau_j'+i\mid p_j\mid
t_j'}\nonumber\\
&&~~~~(\partial_{t_i}^2-\partial_{\tau_i}^2)(\partial_{t_j'}^2-\partial_{\tau_j'}^2)
<T[X(t_1,\tau_1)\cdots X(t_n,\tau_n)X(t_1',\tau_1')\cdots X(t_m',\tau_m')]>,
\nonumber\\
\end{eqnarray}
where Z is the constant for renormalization and is 1 for the theory of no
divergence of loop expansion.

\begin{center}{\bf  III. SYMMETRIC BOUNDARY CONDITION}\end{center}
In this section we will consider the case that $X$-field satisfies the
symmetric
boundary condition.  In this boundary condition, the $X$-field can be written
as
\begin{equation}
X(t,\tau)=\int_0^\infty\frac{dk}{\sqrt{\pi k}}
     [a(k)e^{-i| k| t}+a^\dagger(k)e^{i| k| t}] \cos k\tau.
\end{equation}
To find the S matrix in LSZ reduction formalism, as discussed in previous
section, time-ordered product of $X$-fields  be calculated. For these
calculations, the following formulas will be helpful
\begin{eqnarray}
<T[X(t,\tau) P_{-}(t', \tau')]>&=&
     -2\int_{-\infty}^\infty\frac{dk}{2\pi}\frac{dE}{2\pi}
         (E-k)\cos k\tau  \nonumber\\
&&~~~~~~~~e^{ik\tau'}e^{-iE(t-t')}\Delta^0(E,k),\\
<T[ P_{-}(t,\tau) P_{-}(t', \tau')]>&=&
     \int_{-\infty}^\infty\frac{dk}{2\pi}\frac{dE}{2\pi}
         2ik e^{ik(\tau-\tau')}e^{-iE(t-t')}\Delta(E,k).~~
\end{eqnarray}
In Eqs.(17-18), $\Delta^0$ and $\Delta$ are defined as
\begin{eqnarray}
\Delta^0(E,k) &=& \frac{1}{E^2-k^2+i\epsilon},\\
\Delta(E,k) &=& \frac{1}{E-k+i\mbox{sgn}(k)\epsilon}.
\end{eqnarray}
It is tedious but straightforward applications of Wick's theorem to obtain the
formula
\begin{eqnarray}
&& <T[X(t,\tau) X(t', \tau')]>\nonumber\\
&&= <T[X(t,\tau) X(t', \tau')]>^{(0)}\nonumber\\
& &~~+\frac{i g_{st}^2}{(2\pi)^4}\int_{-\infty}^{\infty}dkdk'dEdk_adk_b
         k_ak_b \cos k\tau \cos k'\tau'e^{-iE(t-t')}\nonumber\\
& &~~~~~~(E-k)g_3(k_a+k_b+k)(E-k')g_3(-k_a-k_b-k')\nonumber\\
& &~~~~~~[\frac{\theta(k_a)\theta(k_b)}{E-k_a-k_b+i\epsilon}
        -\frac{\theta(-k_a)\theta(-k_b)}{E-k_a-k_b-i\epsilon}]
         \Delta^0(E,k)\Delta^0(E,k')\nonumber\\
& &~~+\frac{i g_{st}^2}{2^5\pi^4}\int_{-\infty}^{\infty}dkdk'dEdk_a
         k_a \cos k\tau \cos k'\tau'e^{-iE(t-t')}\nonumber\\
&
&~~~~~~(E-k)(E-k')g_3(k-k'+k_a)g_1(-k_a)\Delta^0(E,k)\Delta(0,k_a)\Delta^0(E,k')
\nonumber\\
& &~~+  O(g_{st}^4),
\end{eqnarray}
where
\[ \theta(x)=\left\{ \begin{array}{ll}
                        1  &\mbox{if $x>0$}\\
                        0  &\mbox{otherwise.}
                     \end{array}
              \right. \]
The second and third term of Eq.(21) come from the one-loop and tadpole
diagrams, respectively. While the superscript $^{(0)}$ denotes that it is
order of $g_{st}^0$, $<T[X(t,\tau) X(t', \tau')]>^{(0)}$  does not contribute
to S matrix.

Though Eq.(21) is rather complicated, S matrix element $<p_{out}\mid p_{in}>$
obtained by applying the LSZ reduction method (Eq.(15)) is simple.
\begin{eqnarray}
&&<p_{out}\mid p_{in}>=\nonumber\\
&&\frac{ig_{st}^2}{2^2\pi}\delta(| p_{in}|-| p_{out}|)
              | p_{in}|\nonumber\\
&&~~\times\int_{-\infty}^\infty dk_a dk_b k_a k_b
[\frac{\theta(k_a)\theta(k_b)}{| p_{in}| +k_a+k_b-i\epsilon}
        -\frac{\theta(-k_a)\theta(-k_b)}{| p_{in}| +k_a+k_b+i\epsilon}]
    \nonumber\\
&&~~~~~~g_3(k_a+k_b+| p_{in}|)g_3(-k_a-k_b-| p_{out}|)\nonumber\\
&&-\frac{ig_{st}^2}{2^5\pi}\frac{\delta(| p_{in}|-| p_{out}|)}
              {| p_{in}|}\int_{-\infty}^{\infty}  dk_a k_a\Delta(0,k_a)
   g_1(-k_a) \nonumber\\
&&~~~\int_{-\infty}^{\infty}dk'
        [\delta(p_{out}  + k')+\delta(p_{out}  - k')](| p_{out}|-k')
           \nonumber\\
&&~~~[(| p_{in}|-p_{in})g_3(k'+k_a-p_{in})+
            (| p_{in}|+p_{in})g_3(k'+k_a+p_{in})]\nonumber\\
&&+O(g_{st}^4)\nonumber\\
&&=\frac{ig_{st}^2}{2^2\pi}\delta(| p_{in}|-| p_{out}|)|
p_{in}|\nonumber\\
&&~~\times\left\{  \begin{array}{l}
\int_{-\infty}^\infty dk_a dk_b k_a k_b
[\frac{\theta(k_a)\theta(k_b)}{| p_{in}| +k_a+k_b-i\epsilon}
        -\frac{\theta(-k_a)\theta(-k_b)}{| p_{in}| +k_a+k_b+i\epsilon}]
    \nonumber\\
{}~~~g_3(k_a+k_b+|  p_{in}| )) g_3(-k_a-k_b-| p_{in}|)\nonumber\\
-\frac{1}{2}\int_{-\infty}^{\infty} dk_a k_a \frac{g_3(k_a)g_1(-k_a)}
                {-k_a+i \mbox{sgn}(k_a)\epsilon} \nonumber\end{array}
\right.\\
& &+O(g_{st}^4)\\
&&=\frac{ig_{st}^2}{24\pi}\delta(| p_{in}|-| p_{out}|)|
p_{in}| \nonumber\\
&&~~\times[i\pi| p_{in}|^3g_3^2(0)+3| p_{in}|^2 \int_{-\infty}^{\infty}
ds g_3(-s)g_3(s)+ \int_{-\infty}^{\infty}ds ~s^2 g_3(-s)g_3(s)\nonumber\\
&&~~~~+3 \int_{-\infty}^{\infty}ds g_1(-s)g_3(s)]
+O(g_{st}^4)\\
&&=\frac{ig_{st}^2}{24}\delta(| p_{in}|-| p_{out}|)|
p_{in}|\nonumber\\
&&~~~\times\left[ \begin{array}{l}
i| p_{in}|^3(\int_{-\infty}^{\infty}f_3(\tau)d\tau)^2
      +6| p_{in}|^2\int_{-\infty}^{\infty}(f_3(\tau))^2d\tau\nonumber\\
+2\int_{-\infty}^{\infty}(\partial_\tau  f_3(\tau))^2d\tau
+6\int_{-\infty}^{\infty}f_3(\tau)f_1(\tau) d\tau\nonumber\end{array}\right. \\
&&+O(g_{st}^4).
\end{eqnarray}
To derive the Eq.(24) from Eq.(23), we use the convolution theorem of Fourier
transforms
\begin{equation}
\int_{-\infty}^{\infty}g_a(s)g_b(-s)ds=2\pi\int_{-\infty}^{\infty}
              f_a(x)f_b(x)dx .
\end{equation}

The above derivation of $<p_{out}\mid p_{in}>$ shows a general property of
S matrix. From the time-ordered product (17), one can find that the following
argument holds for any order of perturbation expansion:
\begin{eqnarray}
&&<X(t,\tau)\cdots >-<X(t,\tau)\cdots >^{(0)}\nonumber\\
&&~~~=\int  dk dE \Delta^0 (E,k) \cos k\tau (E-k)\int dt' e^{-iE(t-t')}\int
d\tau'
 e^{ik \tau'}f_3(\tau')\nonumber\\
&&~~~~~~~~\times\mbox{(terms~ which~do~ not~contain~ $E,k,t,\tau$)}.
\end{eqnarray}
If one consider a LSZ reduction which fixes $k$ to an external momentum $p$,
the contribution to S matrix of the time-ordered product in Eq.(26) is
\begin{eqnarray}
&&\frac{1}{2}\int dkdE[\delta(p+k)+\delta(p-k)]\delta(E\pm| p|)(E-k)
g_3(k+\cdots)\int dt' e^{iEt'}\nonumber\\
&&~~~~~~~~\times\mbox{(terms~ which~do~ not~contain~ $p,E,k$)}\\
&&=\frac{1}{2}\int dk[\delta(p+k)+\delta(p-k)](\mp| p|-k)
g_3(k+\cdots)\int dt' e^{\mp i| p| t'}\nonumber\\
&&~~~~~~~~\times\mbox{(terms~ which~do~ not~contain~ $p,k$)}\nonumber\\
&&=\mp| p| g_3(\pm| p| +\cdots )\int dt' e^{\mp i| p| t'}\nonumber\\
&&~~~~~~~~\times\mbox{(terms~ which~do~ not~contain~ $p$)}.
\end{eqnarray}
Where the upper (lower) sign is for the case that the external energy $| p|$ is
of in- (out-) state. The above argument proves that {\em the S matrix depends
only on the energies of external particles in symmetric boundary condition}.

For the scattering process of three particles, we calculate the time-ordered
product of three $X$-fields:
\begin{eqnarray}
&&<T[X(t_\alpha,\tau_\alpha)X(t_\beta,\tau_\beta)X(t_\gamma,\tau_\gamma)]>
          \nonumber\\
&&~~=\frac{i\sqrt{\pi}g_{st}}{2^3\pi^5}\int_{-\infty}^\infty
dk_\alpha dE_\alpha dk_\beta dE_\beta  dk_\gamma dE_\gamma
 \delta(E_\alpha+E_\beta+E_\gamma)  e^{-iE_\alpha t_\alpha-iE_\beta t_\beta
    -iE\gamma t_\gamma}\nonumber\\
&&~~~~~~\Delta^0(E_\alpha ,k_\alpha  )\Delta^0(E_\beta ,k_\beta
)\Delta^0(E_\gamma ,k_\gamma )
\cos k_\alpha  \tau_\alpha   \cos k_\beta \tau_\beta  \cos k_\gamma \tau_\gamma
 \nonumber\\
&&~~~~~~~(E_\alpha -k_\alpha)(E_\beta  -k_\beta  )(E_\gamma  -k_\gamma  )
g_3(k_\alpha+k_\beta+k_\gamma)\nonumber\\
&&~~~+O(g_{st}^3).
\end{eqnarray}
By applying the formula (15) to this time-ordered product one can find
\begin{eqnarray}
&&<p_\gamma \mbox{out}\mid p_\alpha p_\beta \mbox{in}>\nonumber\\
&&~~~=-g_{st}\delta(| p_\alpha | +| p_\beta |-| p_\gamma |)
\sqrt{| p_\alpha p_\beta p_\gamma|}  g_3(0) +  O(g_{st}^3)\nonumber\\
&&~~~=-g_{st}\delta(| p_\alpha | +| p_\beta |-| p_\gamma |)
\sqrt{| p_\alpha p_\beta p_\gamma|}  \int_{-\infty}^\infty f_3(\tau) d\tau +
O(g_{st}^3).~~~~
\end{eqnarray}

One of the remarkable properties of LSZ reduction is the symmetry between
in- and out-states [15]. If we replace replace energy and momentum ($|
p|$, $p$) of an in-state particle by ($-| p|$, $-p$) besides   square root of
product of external particles' energies in the formula (15), then the formula
gives the S matrix for which  the in-state particle is replaced by a out-state
particle of ($| p|$, $p$). From this symmetry and Eq.(30), one can directly
find, for example, that
\begin{equation}
<p_\beta p_\gamma \mbox{out}\mid p_\alpha\mbox{in}>
=g_{st}\delta(| p_\alpha | -| p_\beta |-| p_\gamma |)
\sqrt{| p_\alpha p_\beta p_\gamma|}  \int_{-\infty}^\infty f_3(\tau) d\tau +
O(g_{st}^3).
\end{equation}

Again, by applying the Wick's theorem, one finds
\begin{eqnarray}
&&<T[X(t_\alpha,\tau_\alpha)X(t_\beta,\tau_\beta)X(t_\gamma,\tau_\gamma)X(t_\delta,\tau_\delta)]>
          \nonumber\\
&&=<T[X(t_\alpha,\tau_\alpha)X(t_\beta,\tau_\beta)X(t_\gamma,\tau_\gamma)X(t_\delta,\tau_\delta)]>^{(0)}
          \nonumber\\
&&~~ -\frac{ig_{st}^2}{2^5\pi^7}\int_{-\infty}^\infty
dk_\alpha dE_\alpha dk_\beta dE_\beta  dk_\gamma dE_\gamma dk_\delta dE_\delta
 \delta(E_\alpha+E_\beta+E_\gamma+E_\delta) \nonumber\\
&&~~~~ e^{-iE_\alpha t_\alpha-iE_\beta t_\beta -iE_\gamma t_\gamma-iE_\delta
t_\delta}  \Delta^0(E_\alpha ,k_\alpha  )\Delta^0(E_\beta ,k_\beta
)\Delta^0(E_\gamma ,k_\gamma )\Delta^0(E_\delta, k_\delta)\nonumber\\
&&~~~~\cos k_\alpha \tau_\alpha  \cos k_\beta \tau_\beta    \cos k_\gamma
\tau_\gamma     \cos k_\delta \tau_\delta
(E_\alpha + k_\alpha  )(E_\beta + k_\beta )(E_\gamma  + k_\gamma   )(E_\delta
+ k_\delta   )\nonumber\\
&&~~~\int_{-\infty}^{\infty} dk k \left[
         \begin{array}{l}
\Delta(E_\alpha+ E_\beta, k)g_3(-k_\alpha-k_\beta+k)g_3(-k_\gamma-k_\delta-k)
        \nonumber\\
+(\beta\leftrightarrow \gamma)+(\beta\leftrightarrow \delta)\nonumber
          \end{array}
         \right.\\
&&~~~~+O(g_{st}^4).
\end{eqnarray}
which, through LSZ reduction method, yields
\begin{eqnarray}
&&<p_\gamma p_\delta\mbox{out}| p_\alpha p_\beta \mbox{in}>\nonumber\\
&&~~~=\frac{i g_{st}^2}{2\pi} \delta(| p_\alpha | +| p_\beta |-| p_\gamma |-|
p_\delta |)
\sqrt{| p_\alpha p_\beta p_\gamma p_\delta|}  \nonumber\\
&&~~~~\times\left[ \begin{array}{l}
   3\int_{-\infty}^\infty   g_3(-s) g_3(s) ds\nonumber\\
    + i\pi g_3^2(0) [| p_\alpha | + | p_\beta |
  +|( | p_\alpha  | - | p_\gamma |)|+|(  | p_\alpha |  -| p_\delta |)|]
\end{array} \right.
\nonumber\\
&&~~~~+O(g_{st}^4)\\
&&~~~=\frac{i g_{st}^2}{2} \delta(| p_\alpha | +| p_\beta |-| p_\gamma |-|
p_\delta |)
\sqrt{| p_\alpha p_\beta p_\gamma p_\delta|}  \nonumber\\
&&~~~~\times\left[ \begin{array}{l}
   6\int_{-\infty}^\infty  (f_3(\tau))^2d\tau \nonumber\\
    + i (\int_{-\infty}^{\infty}f_3(\tau)d\tau)^2 [| p_\alpha | + | p_\beta |
+|( | p_\alpha  | - | p_\gamma |)|+|(  | p_\alpha |  -| p_\delta |)|]
\end{array}
\right.
\nonumber\\
&&~~~~+O(g_{st}^4).
\end{eqnarray}

\begin{center} {\bf IV. ANTISYMMETRIC BOUNDARY CONDITION}\end{center}

For the antisymmetric boundary condition, the $X$-field can be written as
\begin{equation}
X=i\int_0^\infty\frac{dk}{\sqrt{\pi k}}[a(k)e^{-i| k| t}-
   a^\dagger(k)e^{-i| k| t}]\sin k\tau,
\end{equation}
which gives the time-ordered product

\begin{eqnarray}
<T[X(t,\tau) P_{-}(t', \tau')]>&=&
     2i\int_{-\infty}^\infty\frac{dk}{2\pi}\frac{dE}{2\pi}
         (E-k)\sin k\tau  \nonumber\\
&&~~~~~~~~e^{ik\tau'}e^{-iE(t-t')}\Delta^0(E,k).
\end{eqnarray}
Since the above time-ordered product  can be obtained from that of symmetric
boundary condition (Eq.(17)) by replacing $\cos k\tau$  with $\frac{1}{i}\sin
k\tau$, the same
is true for any part of time-ordered product $<T[XX\cdots X]>$ of order of
$g_{st}^n~~(n\geq 1).$ ($<T[XX]>^{(0)}$ in antisymmetric boundary condition
also does not contribute to S matrix as in the symmetric boundary condition.)
As an example,
$<T[X(t_\alpha,\tau_\alpha)X(t_\beta,\tau_\beta)X(t_\gamma,\tau_\gamma)]>$ of
antisymmetric boundary condition can be directly obtained from Eq.(29):
\begin{eqnarray}
&&<T[X(t_\alpha,\tau_\alpha)X(t_\beta,\tau_\beta)X(t_\gamma,\tau_\gamma)]>
          \nonumber\\
&&~~=-\frac{\sqrt{\pi}g_{st}}{2^3\pi^5}\int_{-\infty}^\infty
dk_\alpha dE_\alpha dk_\beta dE_\beta  dk_\gamma dE_\gamma
 \delta(E_\alpha+E_\beta+E_\gamma)  e^{-iE_\alpha t_\alpha-iE_\beta t_\beta
    -iE\gamma t_\gamma}\nonumber\\
&&~~~~~~\Delta^0(E_\alpha ,k_\alpha  )\Delta^0(E_\beta ,k_\beta
)\Delta^0(E_\gamma ,k_\gamma )
\sin k_\alpha  \tau_\alpha   \sin k_\beta \tau_\beta  \sin k_\gamma \tau_\gamma
 \nonumber\\
&&~~~~~~~(E_\alpha -k_\alpha)(E_\beta  -k_\beta  )(E_\gamma  -k_\gamma  )
g_3(k_\alpha+k_\beta+k_\gamma)\nonumber\\
&&~~~+O(g_{st}^3).
\end{eqnarray}

As in the symmetric boundary condition, therefore,
\begin{eqnarray}
&&<X(t,\tau)\cdots >-<X(t,\tau)\cdots >^{(0)}\nonumber\\
&&~~~=\int  dk dE \Delta^0 (E,k) \frac{\sin k\tau }{i}(E-k)\int dt'
e^{-iE(t-t')}\int d\tau' e^{ik \tau'}f_3(\tau')\nonumber\\
&&~~~~~~~~\times\mbox{(terms~ which~do~ not~contain~ $E,k,t,\tau$)}.
\end{eqnarray}
And the contribution to S matrix will be
\begin{eqnarray}
&&\mp \frac{1}{2}\int dkdE[\delta(p+k)-\delta(p-k)]\delta(E\pm| p|)(E-k)
g_3(k+\cdots)\int dt' e^{iEt'}\nonumber\\
&&~~~~~~~~\times\mbox{(terms~ which~do~ not~contain~ $p,E,k$)}\\
&&=\mp \frac{1}{2}\int dk[\delta(p+k)-\delta(p-k)](\mp| p|-k)
g_3(k+\cdots)\int dt' e^{\mp i| p| t'}\nonumber\\
&&~~~~~~~~\mbox{(terms~ which~do~ not~contain~ $p,k$)}\nonumber\\
&&=\mp p g_3(\pm| p| +\cdots )\int dt' e^{\mp i| p| t'}\nonumber\\
&&~~~~~~~~\times\mbox{(terms~ which~do~ not~contain~ $p$)}.
\end{eqnarray}
where the upper(lower) sign is for the case that  the  external momentum  $p$
is of in-(out-)state.

Therefore, {\em the S matrix in antisymmetric boundary condition is
proportional to the momenta of external particles,} while it is proportional to
energies in symmetric boundary condition. Except these parts, S matrix are same
in both cases and described by the energies of external particles.

It is easy to find the following formulas from the results in previous section
and the symmetries between in- and out- states in LSZ reduction method;
\begin{eqnarray}
&<p_{out}\mid p_{in}>&\nonumber\\
&=&\frac{ig_{st}^2}{24}\delta(| p_{in}|-| p_{out}|)
    \frac{p_{in} p_{out}}{| p_{in}|}\nonumber\\
&&~~~\times\left[ \begin{array}{l}
i| p_{in}|^3(\int_{-\infty}^{\infty}f_3(\tau)d\tau)^2
      +6| p_{in}|^2\int_{-\infty}^{\infty}(f_3(\tau))^2d\tau\nonumber\\
+2\int_{-\infty}^{\infty}(\partial_\tau  f_3(\tau))^2d\tau
+6\int_{-\infty}^{\infty}f_3(\tau)f_1(\tau) d\tau\nonumber\end{array}\right. \\
&&+O(g_{st}^4),
\end{eqnarray}
\begin{eqnarray}
&< p_\delta\mbox{out}\mid p_\alpha p_\beta p_\gamma\mbox{in}>&\nonumber\\
&=&-\frac{i g_{st}^2}{2} \delta(| p_\alpha | +| p_\beta |+| p_\gamma |-|
p_\delta |)
\frac{ p_\alpha p_\beta p_\gamma p_\delta}{\sqrt{| p_\alpha p_\beta p_\gamma
p_\delta|}}  \nonumber\\
&&\times\left[ \begin{array}{l}
   6\int_{-\infty}^\infty  (f_3(\tau))^2d\tau \nonumber\\
    + i (\int_{-\infty}^{\infty}f_3(\tau)d\tau)^2 [2| p_\alpha | +
| p_\beta |  +| p_\gamma |+|(  | p_\alpha |  -| p_\delta |)|
						] \end{array}
\right.
\nonumber\\
&&~~~~+O(g_{st}^4).
\end{eqnarray}

\begin{center}{\bf V. APPLICATIONS}\end{center}
For the inverted harmonic oscillator potential $U(\lambda)
=-\frac{\lambda^2}{2}$, $v$ can be written as
\[ v(\tau)=\sqrt{2|\mu|}\cosh \tau.\]
For this case, Eqs. (23), (30), (33) in symmetric boundary condition are
essentially same to those of Ref.[7] except  factors, square root of the
product of external energies, and the results also agree with those in
Ref.[1,2]. So one can easily find S matrices. For examples, in antisymmetric
boundary condition, one can find
\begin{eqnarray}
<p_{out}\mid p_{in}>&=&\frac{ig_{st}^2}{24}\delta(| p_{in}|
    -| p_{out}|)\frac{ p_{in}p_{out}}{| p_{in}|}(i| p_{in}|^3
     +2| p_{in}|^2 +1)\nonumber\\
       &&~~~~~~~~~~~+O(g_{st}^4),
\end{eqnarray}
\begin{eqnarray}
&&<p_\gamma p_\delta\mbox{out}\mid p_\alpha
p_\beta\mbox{in}>=-\frac{g_{st}^2}{2}\delta( | p_\alpha |+|  p_\beta| -|
p_\gamma| -|  p_\delta| )
\frac{ p_\alpha p_\beta p_\gamma p_\delta}{\sqrt{| p_\alpha p_\beta p_\gamma
p_\delta|}}\nonumber\\
&&~~~~~~~~~\times[| p_\alpha |+|  p_\beta|+|( | p_\alpha |-|  p_\gamma| )|+|(|
p_\alpha |-|  p_\delta| )|
   -2i]+O(g_{st}^4).~~~~
\end{eqnarray}
Thanks to the convolution theorem, however, the following formula may be enough
to derive Eqs.(43-44)
\begin{equation}
\int_0^\infty\frac{dx}{\cosh^{2n}(\beta x)}=\frac{2^{2(n-1)}[(n-1)!]^2}
                   {(2n-1)!\beta}.
\end{equation}

As a new application,  we consider a potential
\begin{equation}
U(\lambda) =-\frac{\lambda^2}{2}-\frac{\mu^2}{2\lambda^2}
\end{equation}
with zero fermi energy, which was proposed in Ref.[16] and conjectured to
describe a "naked singularity"[12] (see also Ref.[17]). After some algebras,
one can find that
\begin{eqnarray}
 v(\tau)&=&\sqrt{|\mu|}\frac{\cosh 2\tau}{\sqrt{|\sinh 2\tau|}},
     \\
f_1(\tau)&=&\frac{1}{2|\sinh 2\tau|}+\frac{4|\sinh 2\tau|}{3\cosh^2
    2\tau} -2\frac{|\sinh^3 2\tau |}{\cosh^4 2\tau},\\
\mbox{and}~~~~~~~~~~~&&\nonumber\\
f_3(\tau)&=&\frac{|\sinh 2\tau|}{\cosh^2 2\tau} .
\end{eqnarray}
Making use of general expressions in the preceding  two sections and Eq.(45),
it is easy to find the S matrix elements to the order of $g_{st}^2$.
For example, in symmetric boundary condition,
\begin{eqnarray}
<p_{out}\mid p_{in}>&=&\frac{ig_{st}^2}{24}\delta(| p_{in}|
    -| p_{out}|)| p_{in}|(\frac{i}{9}| p|^3
     +2| p_{in}|^2 +7)\nonumber\\
       &&~~~~~~~~~~~+O(g_{st}^4),
\end{eqnarray}
\begin{equation}
<p_\gamma \mbox{out}\mid p_\alpha p_\beta\mbox{in}>=
    -\frac{g_{st}}{3}\sqrt{| p_\alpha p_\beta p_\gamma |}
\delta( | p_\alpha |+|  p_\beta|  -|  p_\gamma| )
+O(g_{st}^3).~~~~~~~~~~
\end{equation}
\begin{eqnarray}
&&<p_\gamma p_\delta\mbox{out}\mid p_\alpha p_\beta\mbox{in}>
=-\frac{g_{st}^2}{18}\delta( | p_\alpha |+|  p_\beta| -|
  p_\gamma| -|  p_\delta| )\sqrt{| p_\alpha p_\beta p_\gamma
   p_\delta|}\nonumber\\
&&~~~~~~~\times[| p_\alpha |+|  p_\beta|+|( | p_\alpha |-|  p_\gamma| )|+|(|
p_\alpha |-|  p_\delta| )|
   -18i]+O(g_{st}^4).~~~~~~
\end{eqnarray}
The fact that $<p_\gamma \mbox{out}| p_\alpha p_\beta\mbox{in}>\neq 0$ has
been noted in Ref.[12].

\begin{center}{\bf VI. CONCLUSION  }\end{center}

We have studied the S matrix of $\mu<0$ collective field theory by applying the
LSZ reduction method. We consider the spatially symmetric and antisymmetric
boundary conditions. In the antisymmetric boundary condition, the S matrices
are proportional to the momenta of external particles, while in symmetric
boundary condition they are proportional to the energies. Besides these
factors, the S matrices are same in
both boundary conditions. In the sense that the S matrices are described by
natural variables (momenta and/or energies), both of the boundary conditions
may
be thought to be physically acceptable.

To the order of $g_{st}^2$, we find simple formulas of S matrices in terms of
velocity. In Ref.[12] (see also Ref.[17]), it has been hypothesized that for
the theory of black hole the S matrix of odd particles must vanish.
This is not possible in  $\mu<0$ case since the S matrix of three external
particles proportional to $\int_{-\infty}^\infty\frac{1}{v^2(\tau)}d\tau$ and
$v$ is always larger than 0. Furthermore, it mostly looks like that the
$\mu<0~$ S matrices of different potentials have universal form  when they are
expanded in terms of energies of external particles. The effects of different
potentials would appear only in the coefficients of the expansions.

\vspace*{2cm}
\begin{center}{\bf ACKNOWLEDGMENT}\end{center}
The author thanks Professor A. Jevicki for advices and comments. This work was
supported in part by the Korea Science and Engineering Foundation, and also by
the U.S. Department of Energy under contract No. DE-AC02-76-ER03130.
\newpage

\noindent
[1]G. Moore, Nucl. Phys. B368, 557 (1992).\newline
[2]G. Moore, R. Plesser and S. Ramgoolam, Nucl. Phys. B377, 143 (1992).\newline
[3]S.R. Das and A. Jevicki, Mod. Phys. Lett. A5, 1639 (1990).\newline
[4]J. Polchinski, Nucl. Phys. B362, 125 (1991).\newline
[5]G. Moore and R. Plesser, Phys. Rev. D46, 1730 (1992).\newline
[6]K. Demeterfi, A. Jevicki and J.P. Rodrigues, Nucl. Phys. B362, 173 (1991);
{\em ibid.} B365, 499 (1991).\newline
[7]I.R. Klebanov, in {\em String Theory and Quantum Gravity}, edited by J.
Harvey et al.  (World Scientific, Singapore, 1992);E. Hsu and I.R. Klebanov,
Phys. Lett. B321,99 (1994).\newline
[8]G. Mandal, A. Sengupta and S. Wadia, Mod. Phys. Lett. A6, 1465
(1991).\newline
[9]R. Jackiw, in {\em Quantum Theory of Gravity} edited by S. Christensen
(Hilger, Bristol, 1984).\newline
[10]D.Y. Song, Phys. Rev. D49, 6794 (1994); {\em ibid.} D48, 3925
(1993).\newline
[11]H. Lehmann, K. Symanzik and W. Zimmermann, Nuovo Cimento, 1, 205 (1955);
for review, see J.D. Bjorken and S.D. Drell, {\em Relativistic Quantum Fields},
(McGraw-Hill, New York, 1965).\newline
[12]A. Jevicki and T. Yoneya, Nucl. Phys. B411, 64 (1994).\newline
[13]K. Demeterfi, I.R. Klebanov and J.P. Rodrigues,  Phys. Rev. Lett. 71, 3409
(1993).\newline
[14]D.J. Gross and I.R. Klebanov, Nucl. Phys. B359, 3 (1991).\newline
[15]For example, see C. Itzykson and J.-B. Zuber, {\em Quantum Field Theory},
(McGraw-Hill, New York, 1980).\newline
[16]Z. Yang, University of Rochester  report, UR-1251/hepth-9202078
(unpublished).\newline
[17]M. Natsuume and J. Polchinski, Nucl. Phys. B424, 137
(1994)/hepth-9402156.\newline

\end{document}